RESEARCH ARTICLE

# Experts' Cognition-Driven Ensemble Deep Learning for External Validation of Predicting Pathological Complete Response to Neoadjuvant Chemotherapy from Histological Images in Breast Cancer

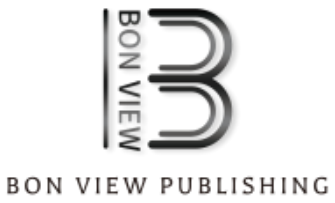

**Yongquan Yang[1,2,†,*], Fengling Li[1,3,†], Yani Wei[1,3], Yuanyuan Zhao[4], Jing Fu[5], Xiuli Xiao[6], Hong Bu[1,3,*]**

*1 Institute of Clinical Pathology, West China Hospital, Sichuan University, China.*
*2 Zhongjiu Flash Medical Technology Co., Ltd., China.*
*3 Department of Pathology, West China Hospital, Sichuan University, China.*
*4 Department of Pathology, Shanxi Provincial Cancer Hospital / Shanxi Hospital Affifiliated to Cancer Hospital, Chinese Academy of Medical Sciences / Cancer Hospital Affifiliated to Shanxi Medical University, China.*
*5 Department of Pathology, Sichuan Provincial People's Hospital, China.*
*6 Department of Pathology, The Affiliated Hospital of Southwest Medical University, China.*

**†Co-first author**
***Corresponding author:** Yongquan Yang, Institute of Clinical Pathology, West China Hospital, Sichuan University and Zhongjiu Flash Medical Technology Co., Ltd., China. Email: The research was conducted at Institute of Clinical Pathology, West China Hospital, Sichuan University and Zhongjiu Flash Medical Technology Co., Ltd., China. Correspondence concerning this article should be addressed to remy_yang@foxmail.com and Hong Bu, Department of Pathology and Institute of Clinical Pathology, West China Hospital, Sichuan University, China. Email: hongbu@scu.edu.cn.

**Abstract:** In breast cancer, neoadjuvant chemotherapy (NAC) provides a standard treatment option for patients who have locally advanced cancer and some large operable tumors. A patient will have better prognosis when he has achieved a pathological complete response (pCR) with the treatment of NAC. There has been a trend to directly predict pCR to NAC from histological images based on deep learning (DL). However, the DL-based predictive models numerically have better performances in internal validation than in external validation. In this paper, we aim to alleviate this situation with an intrinsic approach. We propose an experts' cognition-driven ensemble deep learning (ECDEDL) approach. Taking the cognition of both pathology and artificial intelligence experts into consideration to improve the generalization of the predictive model to the external validation, ECDEDL can intrinsically approximate the working paradigm of a human being which will refer to his various working experiences to make decisions. ECDEDL was validated with 695 WSIs collected from the same center as the primary dataset to develop the predictive model and perform the internal validation, and was also validated with 340 WSIs collected from other three centers as the external dataset to perform the external validation. In external validation, ECDEDL improves the AUCs of pCR prediction from 61.52(59.80-63.26) to 67.75(66.74-68.80) and the Accuracies of pCR prediction from 56.09(49.39-62.79) to 71.01(69.44-72.58). ECDEDL was quite effective for external validation of predicting pCR to NAC from histological images in breast cancer, numerically approximating the internal validation.

## 1. Introduction

With the advances of deep learning (DL) [1], mostly deep neural networks [2-4] which are the state-of-the-art machine learning techniques, various studies have shown that DL-based artificial intelligence (AI) models have significant effectiveness and potentials in medical diagnostic or prognostic prediction [5-8]. There has been an increasingly standardized paradigm of constructing DL-based AI models for medical diagnostic or prognostic prediction: Primarily, clinical data and corresponding diagnostic and prognostic results are collected as a training dataset; Subsequently, a deep learning architecture is selected and optimized on the collected training dataset to produce a DL-based AI model that can predict the diagnostic or prognostic results corresponding to the clinical data; Finally, the produced DL-based AI model is validated on some new data that contain clinical data and corresponding diagnostic and prognostic results unseen in the training dataset.

Particularly, the validation procedure is an essential part in the constructing paradigm, for it can reflect the expected predictive performance and generalization of the produced DL-based AI model in practical usage. The data required by the validation procedure can be internal or external. Specifically, the internal data is assumed to have the same distribution while the external data is assumed to have a different distribution, compared with the training data collected for producing the DL-based AI model. Usually, we say that the data for the validation procedure is internal and have the same distribution compared with the training data when the data for the validation procedure and the training data are collected from the same centers, since they share common data production in the same centers. On the contrary, we say that the data for the validation procedure is external and have a different distribution compared with the training data when the data for the validation procedure and the training data are collected from different centers, since they probably have uncommon data production in different centers. The validation procedure is called internal validation when provided with internal data, or external validation when provided with external data. Both internal validation and external validation are essential [9-13], since internal validation can reflect the feasibility of constructing a DL-based AI model for a medical diagnostic or prognostic prediction task while external validation can reflect the potentials of the constructed DL-based AI model for a wider usage in practice.

In breast cancer, neoadjuvant chemotherapy (NAC) [14, 15] provides a standard treatment option for patients who have locally advanced cancer and some large operable tumors. In clinical trials, it has been shown that a patient will have better prognosis when he has achieved a pathological complete response (pCR) with the treatment of NAC to reduce the tumor burden and promote breast-conserving surgery [16]. In breast cancer imaging, there has been a trend to directly predict pCR to NAC from histological images using DL [17]. Following the paradigm of constructing DL-based AI models for medical prediction, which has been described in the first paragraph of this section, existing studies [18-22] have provided alternative solutions for predicting pCR from histological images in breast cancer. The application of these alternative solutions can be summarized as directly build the predictive model from the histological images via DL.

However, it has been a commonly known problem that the AI models constructed for medical prediction numerically have better performances in internal validation than in external validation, which significantly affects the clinical safety of using AI models [23]. The primary reason for this situation lies in that the distribution of the external data for validation is different from the distribution of the training data for the construction of the predictive model, due to the significant variance of slide preparation and microscope scanning. This issue is also known as sample selection bias [24, 25] in statistics with small data, or out-of-distribution validation [26] and domain adaption/generalization [27-30] in machine learning with big data. According to a recent survey [31], the out-of-distribution problem of external validation, i.e., the problem of domain generalization is the primary challenge for deep learning in breast cancer imaging. Therefore, it is important and necessary to investigate advanced approaches for the out-of-distribution problem of external validation in breast cancer imaging.

The usual methods for alleviating this out-of-distribution problem of external validation in breast cancer imaging can be divided into three categories [31], including data augmentation via color distortion [32, 33], domain adaption from source domain to target domain via adversarial learning [34, 35], and domain generalization via feature alignment and domain-invariant feature learning [36-38]. Having shown promising potentials to provide alternative solutions for external validation of predicting pCR to NAC in breast cancer from histological images, these existing usual methods for as well have some limitations. Data augmentation via color distortion usually can only imitate certain aspects of variance due to the complex situation of the slide preparation and microscope scanning. Domain adaption via adversarial learning requires some image samples of the target domain in advance, which is not suitable for the situation where the target domain is unseen. Domain generalization via feature alignment and domain-invariant feature learning does not require some image samples of the target domain in advance, however, only a very few methods have been particularly proposed for tasks in medical analysis [39-42]. These approaches can be summarized as addressing the problem via making the internal data and external data less different, without targeting at the pattern that probably will not change underlying both the internal data and the external data.

In addition to these three categories of usual methods, two recent works [43, 44] have proposed to employ federated learning [45-47] to improve performance in multicenter deep learning without data sharing, which have shown that federated learning can help to provide alternative solutions relevant to the out-of-distribution problem of external validation in medical prediction. However, a federated learning solution requires each of the multiple centers constructs a predictive model, and it also needs a central system to manage communications between the predictive models of the multiple centers for testing in practical usage. As a result, federate leaning is a technique that is more appropriate to solve the problem of data privacy among multiple centers and requires high expenses at the meantime. Thus, federated learning is theoretically less intrinsic for external validation via requiring more managing and computing resources to be implemented.

We observe that these existing alternative methods have the paradigm that is different from the working paradigm of a medical expert. Usually, a pathological expert commonly will refer to his cognition, that has been accumulated via different working experiences, about the medical data at hand to make decisions. The out-of-distribution problem of the external data commonly will not affect a well-trained expert that much to make usual decisions, since a first-class expert is unlikely to become a third-class expert in practice because of the distribution change of data. Due to these insights, we argue that an intrinsic approach needs to be proposed for addressing the out-of-distribution problem in external validation.

In this paper, we propose an experts' cognition-driven ensemble deep learning (ECDEDL) approach for external validation of predicting pCR to NAC from histological images in breast cancer. The proposed ECDEDL, which takes the cognition of both pathology and AI experts into consideration to improve the generalization of the predictive model to the external validation, has three innovations: 1) Proposing a data preparation strategy that takes into account the cognition of pathology experts about viewing a histological image in breast cancer, which results in a Tumor dataset and a Stroma dataset respectively extracted from histological images; 2) Proposing a learning paradigm that takes into account the cognition of AI experts about exploiting complementary information among the pathologically variant contents of histological images to improve the generalization of the predictive model, which results in an Ensemble Deep Learning [48] framework; 3) Constructing a new approach for the external validation of predicting pCR to NAC from histological images in breast cancer, by integrating the proposed pathology experts' cognition-driven data preparation strategy and the proposed AI experts' cognition-driven learning paradigm. Regarding to these three innovations, the proposed ECDEDL approach is different from the existing approaches [39-44] for the out-of-distribution problem of external validation in medical prediction.

Since the cognition of both pathology and AI experts have been taken to construct the predictive model, the proposed ECDEDL approach, to some extent, approximates the working paradigm of a human being which will refer to his various working experiences to make decisions. As the cognition of experts is less likely to be affected by data distribution shift in practice, the proposed ECDEDL approach for external validation is more likely to be invariant to the significant variance of slide preparation and microscope scanning of the external data.

As far as we know, this paper is the first that has proposed an experts' cognition-driven approach, which is particularly for addressing the out-of-distribution problem in external validation. On the basis of the task of predicting pCR to NAC from histological images in breast cancer, the contributions of this work include: 1) Proposing a novel ECDEDL approach for external validation; 2) Implementing and evaluating the proposed ECDEDL approach for improving the performance for external validation; 3) The proposed ECDEDL approach shows significant effectiveness in improving the performance for external validation.

## 2. Materials and Methods

This section is structured as follows: Primarily in Section 2.1, we give descriptions for the components of the proposed ECDEDL approach; Subsequently in Section 2.2, we present specific implementation details of ECDEDL; Finally in Section 2.3, we discuss the evaluating strategy of ECDEDL for external validation.

### 2.1. Proposed ECDEDL

ECDEDL constitutes of a pathology experts' cognition-driven data preparation strategy, an AI experts' cognition-driven learning paradigm, and the feeding relation between them. The methodology of ECDEDL for external validation of predicting pCR to NAC from histological images in breast cancer is shown as Figure 1.

#### *2.1.1. Pathology experts' cognition-driven data preparation strategy*

We propose a data preparation strategy which respectively extracts a Tumor dataset and a Stroma dataset from the collected histological images. This data preparation strategy is established by referring to the cognition of pathology experts about viewing a histological image in breast cancer, which is that tumor and stroma are likely to be paid more attention than the whole content of the histological image and can probably possess potential predictive ability for predicting pCR to NAC from histological images in breast cancer. Moreover, two recent studies [49, 50] have shown that tumor and stroma areas of histological images in breast cancer can both be predictive for pCR to NAC. Proving that the cognition of pathology experts indeed can have effectiveness in constructing appropriate DL-based AI models for medical prediction, these two studies [49, 50] just can explain the rationality of the proposed pathology experts' cognition-driven data preparation strategy for the external validation of predicting pCR to NAC from histological images in breast cancer.

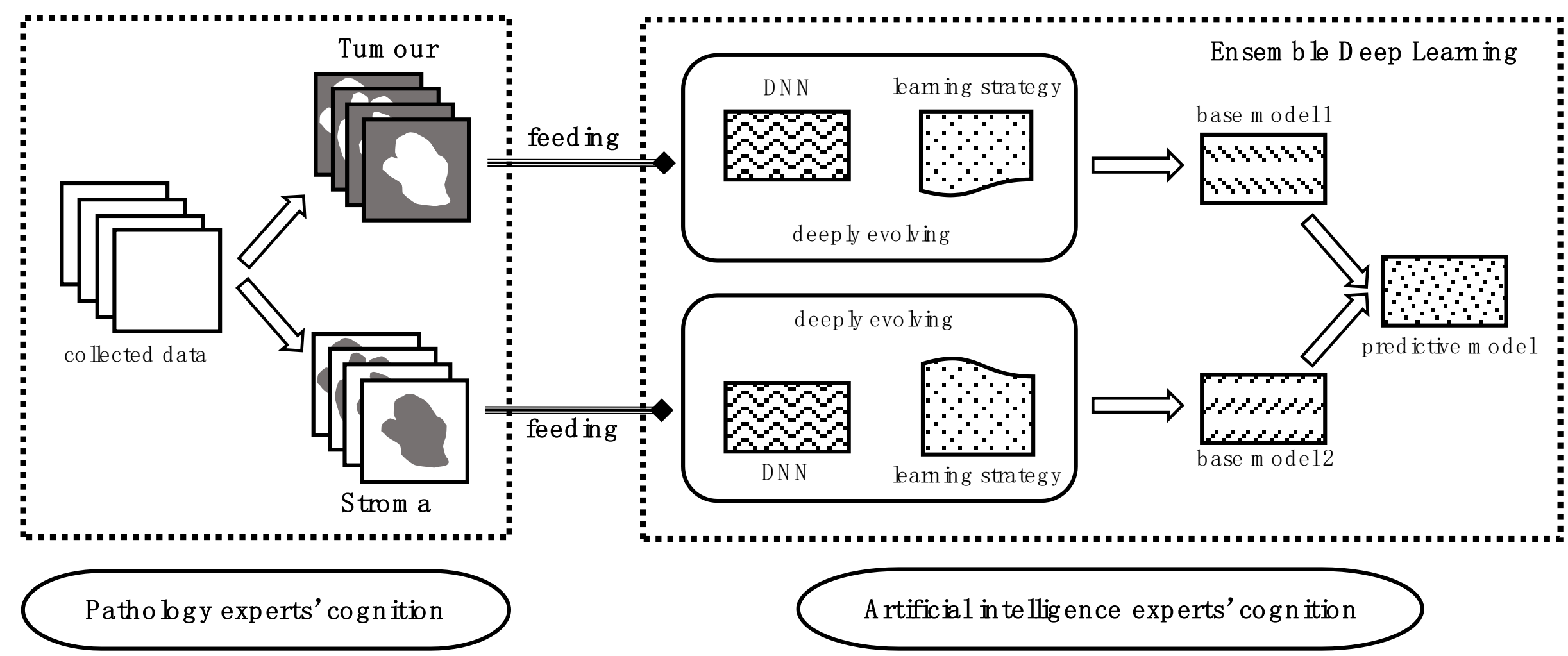

**Figure 1. Outline of the proposed ECDEDL approach for external validation of predicting pCR to NAC from histological images in breast cancer.**

### *2.1.2. Artificial intelligence experts' cognition-driven learning paradigm*

We propose an artificial intelligence (AI) experts' cognition-driven learning paradigm which is an Ensemble Deep Learning framework. This learning paradigm is established by referring to the cognition of AI experts about exploiting complementary information among the pathologically variant contents of histological images to improve the generalization of the predictive model, which can be realized via ensemble learning [48, 51, 52] based on data manipulation, which is an effective basis for the realization of ensemble learning. Moreover, some researches [29, 30, 53-55] have shown that ensemble deep learning have the potentials in addressing the out-of-distribution problem of external validation. Proving that the cognition of AI experts can also have effectiveness in constructing better DL-based AI models, these researches [29, 30, 53-55] just can explain the rationality of the proposed AI experts' cognition-driven learning paradigm for the external validation of predicting pCR to NAC from histological images in breast cancer.

### *2.1.3. Feeding relation for construction of ECDEDL*

Based on the pathology experts' cognition driven data preparation strategy proposed in section 2.1.1 and the AI experts' cognition driven learning paradigm proposed in section 2.1.2, we constructed ECDEDL by feeding the Tumor and Stroma datasets of the data preparation strategy to Ensemble Deep Learning framework of the learning paradigm as the data manipulation basis, as the Tumor and Stroma datasets of the data preparation strategy naturally fit the data manipulation basis for the ensemble deep learning framework of the learning paradigm.

## 2.2. Implementation details of ECDEDL

### *2.2.1. Data basis and preprocessing*

The histological images used in this study to evaluate ECDEDL for external validation are the same as our previous paper [50], in which more details are provided. The used histological images were 1035 whole slide images (WSIs) collected from four centers. Among the 1035 WSIs, 695 WSIs collected from the same center are used as the primary dataset to develop the predictive model and perform the internal validation, and the rest 340 WSIs collected from other three centers are used as the external dataset to perform the external validation. More details are available at F. Li et al. in 2022 [50]. Among the primary dataset, 555 WSIs are used as the training dataset to develop the predictive model and the rest 140 WSIs are used as the internal dataset to perform the internal validation.

As a WSI usually contains many repetitive and less informative regions, pathological experts were invited to annotate representative regions containing tumor and stroma on each of the collected WSIs. The annotated representative region is called region of interest (ROI), which ensures that the stroma inside the ROI was near the tumor and surrounded by tumor cells. Small images from the ROIs annotated on each of the collected WSIs were cropped at 233 × 233 μm squares (256 × 256 pixels at 10 × magnification), which are called "tiles". More details about how the ROIs were annotated, readers can refer to F. Li et al. in 2022 [50].

The data basis and preprocessing can be summarized as Table 1.

**Table 1. Summarization of data basis and preprocessing.**

| Total 1035 WSIs | | | |
|---|---|---|---|
| Primary dataset (695 WSIs from the same center) | | | External dataset (340WSIs from other 3 centers) |
| Training dataset (555WSIs) | | Internal dataset (140 WSIs) | Total (18304 tiles) |
| Total (32556 tiles) | | Total (6981 tiles) | |
| Training (26045 tiles) | Validation (6511 tiles) | | |
| Per WSI (59 tiles on average) | | Per WSI (50 tiles on average) | Per WSI (54 tiles on average) |
| Model development | | Internal validation | External validation |

### *2.2.2. Technical details*

For the preparation of the Tumor and Stroma datasets, we used a previously developed image segmentation tool to extract the Tumor and the Stroma contents from the collected histological images tiles. Readers can refer to Yang et al. in 2024 [56], for the technical details of the used image segmentation tool. Some examples of the Tumor and Stroma datasets extracted from the original histological image tiles are shown as Figure 2.

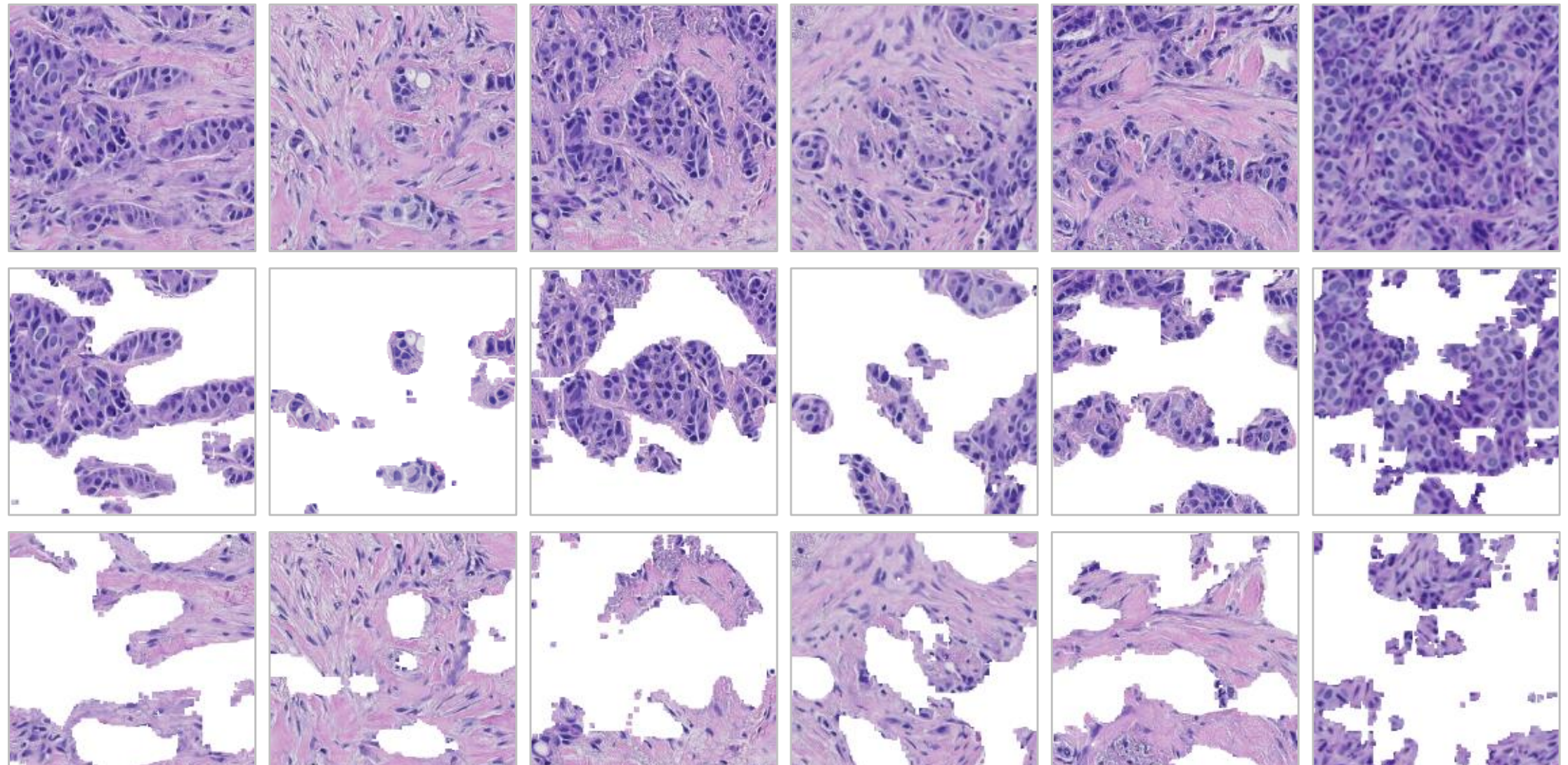

**Figure 2. Examples of original histological image tiles and extracted Tumor and Stroma tiles. Top: original histological image tiles; Middle: extracted Tumor tiles; Bottom: extracted Stroma tiles.**

For the implementation of the Ensemble Deep Learning framework, two key points need to be considered: 1) the settings of DNN architecture and learning strategy for generating base models; and 2) the ensembling criterion for forming the final predictive model. For point 1), we employ existing state-of-the-art deep convolutional neural networks (DNNs) [57] and corresponding learning strategy to generate base models. More specifically, we respectively employed MobileNetV2 [58], ResNet101V2 [59] and NASNetLarge [60] as the DNN architecture from light weight to complex, which will be further discussed in the next section of evaluating ECDEDL for external validation.

The details of the learning strategy include, optimizer: SGD [61] with learning rate=0.001, momentum=0.9; batch size: 16; epochs: 256; online augmentation: horizontal flip=True, vertical flip=True, rotation range=10, zoom range=[0.8, 1.2], width shift range=0.2, height shift range=0.2, brightness range=[0.7, 1.3]; and weighted cross-entropy loss. For point 2) we employ weighted average strategy to integrate the predictions of the base models for forming the final predictive model. More specifically, we weighted the base models according to their individual predictive performance (default is fifty-fifty).

For the evolvement of ECDEDL, we firstly used the corresponding learning strategy to optimize the DNN architecture to produce two base models, respectively feeding the prepared Tumor and Stroma datasets to the DNN architecture. Then, based on the produced two base models, we weighted and averaged their predictions to form the final prediction.

## 2.3. Evaluating strategy of ECDEDL for external validation

We respectively trained four series of predictive models for predicting pathological complete response to neoadjuvant chemotherapy from histological images in breast cancer. The trained four series of predictive models include: 1) the predictive model produced by training the given DNN architecture based on the original prepared data, examples of which are shown as the top row of Figure 2; 2) the predictive model produced by training the same DNN architecture based on the Tumor dataset, examples of which are shown as the middle row of Figure 2; 3) the predictive model produced by training the same DNN architecture based on the Stroma dataset, examples of which are shown as the bottom row of Figure 2; 4) the ensemble model of the predictive models respectively produced based on the Tumor and Stroma dataset, which represents the ECDEDL approach.

The four series of predictive models were trained with the same learning strategy described in section 2.2.2. We respectively denote the four series of predictive model as Direct model, Tumor model, Stroma model and TS-Ensemble model. We validate and compare their performances using different metrics, to show the effectiveness of the ECDEDL approach for external validation. More specifically, we first compare the Tumor and Stroma models with the Direct model to show the effectiveness of the pathology experts' cognition in ECDEDL. Second, we compare TS-Ensemble model respectively with Tumor model and Stroma model to assess the effectiveness of the AI experts' cognition in ECDEDL. Then, we compare TS-Ensemble model with the Direct model to show the effectiveness of the ECDEDL approach for external validation, since the Direct model can be regarded as the usual model constructed without experts' cognition. To avoid the effects of possible experimental bias errors, we repeated the training and validation of the four series of predictive models five times and summarize corresponding evaluation metrics for a fair comparison.

We respectively employed MobileNetV2 [58], ResNet101V2 [59] and NASNetLarge [60] (from light weight to complex) as the DNN architecture in the procedures of the training and validation of the two series of predictive models, to show the stability of the effectiveness of the ECDEDL approach for external validation with different DNN architectures. Particularly, from light weight to complex DNN architectures are chosen for experiments, in order to simulate the situation of real-world applications regarding to the consideration of computing resource and efficiency. We employed ROC and PR curves to evaluate the overall performances of predictive models, and metrics of Precision, Recall, F1 and Accuracy calculated at the threshold of probability 0.5 evaluate the practical performances of predictive models, as well as different probabilities from 0 to 1 to evaluate the ablation performances of predictive models.

## 3. Results and Discussion

In this section, we show the effectiveness of the proposed ECDEDL referring to the evaluating strategy in Section 2.3. The results and discussion are enclosed in the experiments regarding the training dataset, the internal dataset and the external dataset in Table 1. Primarily, difference predictive models were produced based on the training dataset. Subsequently, the predictive models were respectively evaluated on the internal dataset and the external dataset. Finally, the evaluated results were discussed to show the effectiveness of the proposed ECDEDL for external validation.

Basically, it is reasonable to assume the internal dataset is independent and identically distributed (i.i.d) with the training dataset, since they are from the same primary dataset. However, we assume the external dataset is independent and non-identically distributed with the primary dataset (both the training dataset and the internal dataset), since it is independently and differently collected regarding the primary dataset. Thus, a predictive model produced based on the training dataset will have better performance on the internal dataset than the external dataset. And, the performance of a predictive model on the external dataset closer to the performance on the internal dataset is better.

Notably, as the external dataset was collected with limited number (340) of WSIs, the distribution of the external dataset can inevitably be not able to fully represent the distribution of the real-world external dataset beyond the primary dataset.

## 3.1. Effectiveness of experts' cognition in ECDEDL

The ROC and PR curves of the Tumor model, the Stroma model and the Direct model on external data are shown as Figure 3-A. In Figure 3-A, the curves were drawn with different DNN architectures and repeated experiments. From Figure 3-A, we can summarize that the Tumor model and the Stroma model perform better than both the Direct model on external data, which reflects that pathology experts' cognition in ECDEDL is significantly effective, leading to better performances for external validation.

The ROC and PR curves of the TS-Ensemble model (ensemble of the Tumor and Stroma models), the single Tumor model and the single Stroma model on external data are shown as Figure 3-B. In Figure 3-B the curves were drawn with different DNN architectures and repeated experiments. From Figure 3-B, we can summarize that the TS-Ensemble model performs better than both the Tumor model and the Stroma model on external data, which reflects that AI experts' cognition in ECDEDL is comparatively effective, leading to even better performances for external validation.

## 3.2 Overall performance of ECDEDL

The ROC curves of the TS-Ensemble model on external data and the Direct model respectively on Internal data and external are shown as Figure 4-A. In Figure 4-A, the curves were drawn regarding to different DNN architectures with repeated experiments, and their union results. The corresponding PR curves are shown as Figure 4-B. The 95% confident intervals (CI) of the AUCs for ROC curves corresponding to Figure. 4-A and the APs for PR curves corresponding to Figure 4-B are shown as Table 2. Form Figure 4 and Table 2, we can summarize that the TS-Ensemble model performs much better than the Direct model on external data, and the performances of the TS-Ensemble model on external data are close to the performances of the Direct model on internal data. These results indicate that the overall performances of the ECDEDL approach for external validation are quite effective, approximating the internal validation.

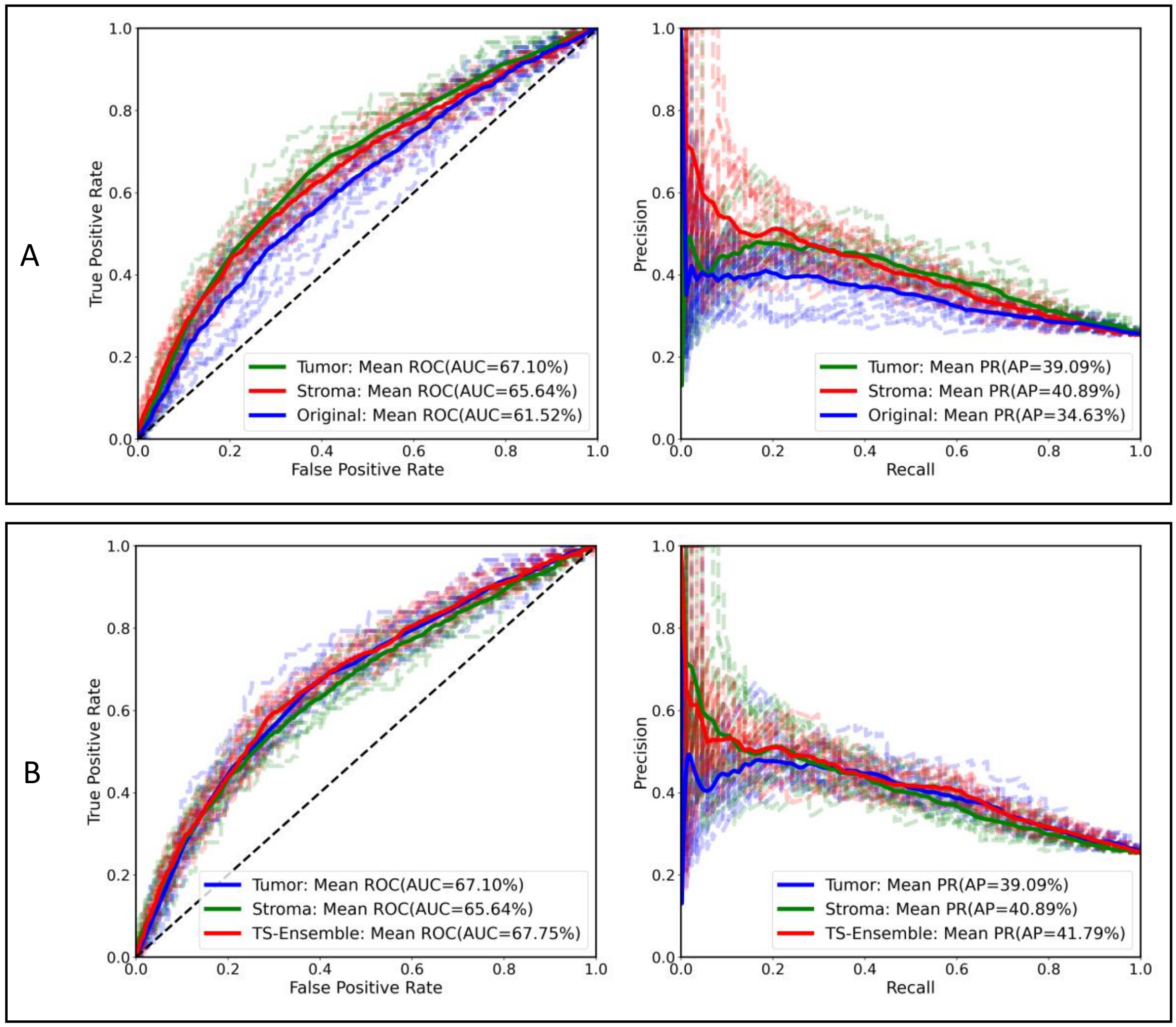

**Figure 3. ROC and PR curves of various models on external data. A: The ROC and PR curves of the Tumor model, the Stroma model and the Direct (original) model on external data. B: The ROC and PR curves of the TS-Ensemble model (ensemble of the Tumor and Stroma models), the single Tumor model and the single Stroma model on external data.**

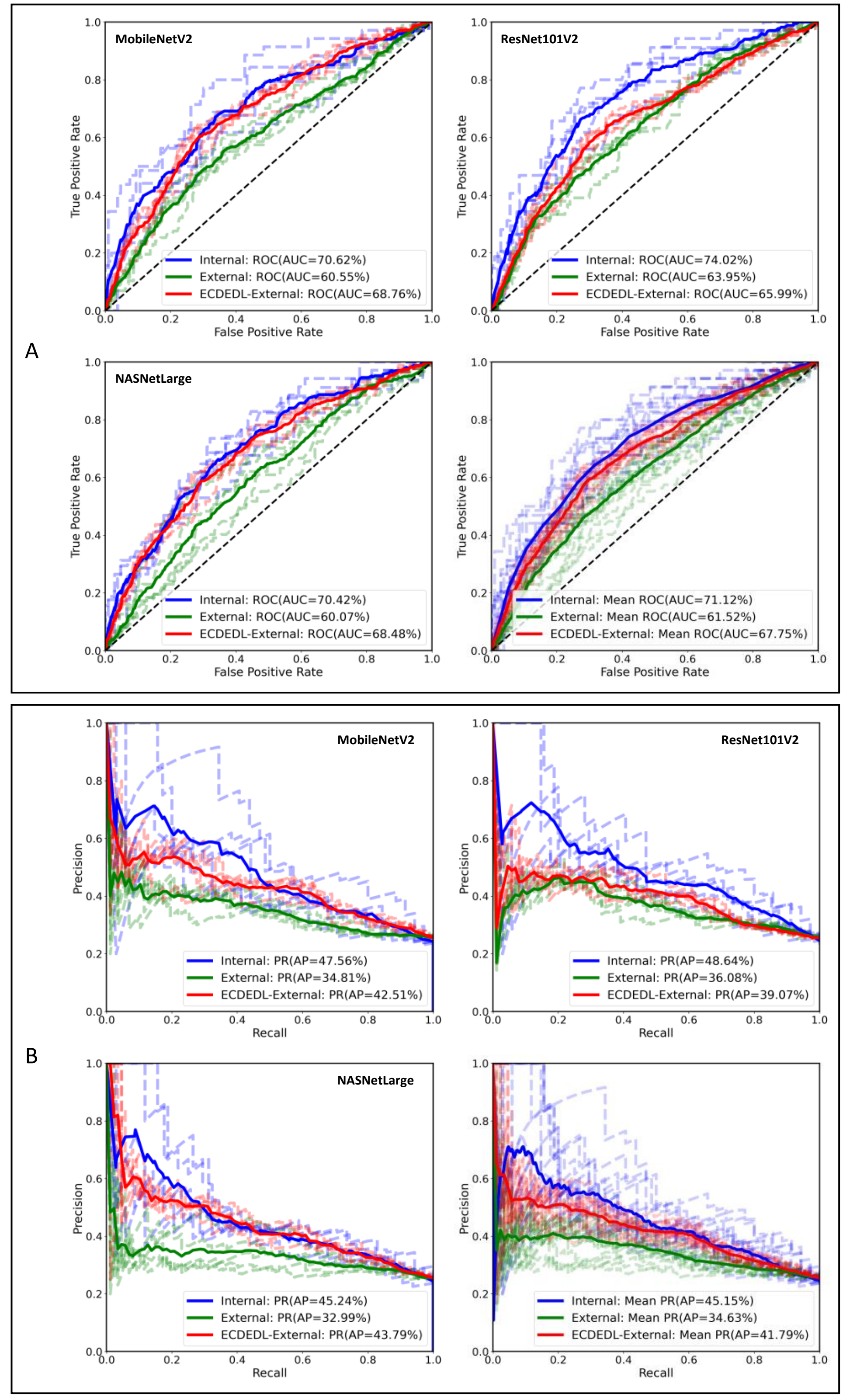

**Figure 4. ROC curves of various models on internal dataset and external dataset. Internal: Direct model on internal data; External: Direct model on external data; ECDEDL-External: TS-Ensemble model on external data. A: ROC curves. B: PR curves.**

**Tabel 2. The 95% confident intervals (CI) of the AUCs for ROC curves and the Aps for PR curves corresponding to Figure 4**

| Metrics-Architecture | | Internal(CI) | External(CI) | ECDEDL-External(CI) |
|---|---|---|---|---|
| AUC | MobileNetV2 | 70.62(66.12-75.07) | 60.55(58.57-62.53) | 68.76(66.90-70.67) |
| | ResNet101V2 | 74.02(70.87-77.11) | 63.95(62.30-65.59) | 65.99(65.15-66.88) |
| | NASNetLarge | 70.42(68.69-72.16) | 60.07(56.06-64.12) | 68.48(66.97-70.07) |
| | Union | 71.12(69.71-73.64) | 61.52(59.80-63.26) | 67.75(66.74-68.80) |
| AP | MobileNetV2 | 47.56(38.95-53.76) | 34.81(32.71-36.13) | 42.51(40.31-43.72) |
| | ResNet101V2 | 48.63(42.00-52.39) | 36.08(34.14-37.23) | 39.07(37.39-39.96) |
| | NASNetLarge | 45.24(40.70-48.12) | 32.99(28.73-36.22) | 43.79(41.94-44.55) |
| | Union | 45.15(42.95-49.21) | 34.63(32.67-35.72) | 41.79(40.04-42.58) |

## 3.3. Practical performance of ECDEDL

The 95% confident intervals (CI) for the metrics of Precision, Recall, F1 and Accuracy at the threshold of probability 0.5 for the TS-Ensemble model on external data and the Direct model respectively on internal data and external data are shown as Table 3.

**Table 3. The 95% confident intervals (CI) for the metrics of Precision, Recall, F1 and Accuracy of the TS-Ensemble model on external data and the Direct model respectively on internal data and external data. Internal: Direct model on internal data; External: Direct model on external data; ECDEDL-External: TS-Ensemble model on external data.**

| Metrics-Architecture | | Internal(CI) | External(CI) | ECDEDL-External(CI) |
|---|---|---|---|---|
| Precision | MobileNetV2 | 61.07(46.28-75.87) | 40.32(37.59-43.05) | 45.98(41.18-50.78) |
| | ResNet101V2 | 47.24(43.24-51.25) | 32.40(29.18-35.62) | 40.87(37.66-44.08) |
| | NASNetLarge | 41.88(40.64-43.12) | 29.05(26.60-31.50) | 45.14(42.11-48.18) |
| | Union | 50.07(43.70-56.43) | 33.92(31.03-36.82) | 44.00(41.67-46.33) |
| Recall | MobileNetV2 | 30.74(26.41-35.07) | 29.43(16.18-42.67) | 33.10(20.06-46.15) |
| | ResNet101V2 | 45.48(34.13-56.82) | 70.80(56.69-84.92) | 51.72(42.25-61.20) |
| | NASNetLarge | 50.82(44.02-57.62) | 78.62(65.47-91.77) | 42.30(34.75-49.85) |
| | Union | 42.34(36.16-48.53) | 59.62(46.20-73.03) | 42.38(35.59-49.16) |
| F1 | MobileNetV2 | 40.36(35.34-45.38) | 31.55(20.96-42.14) | 37.04(28.82-45.25) |
| | ResNet101V2 | 45.77(38.03-53.50) | 43.40(41.99-44.82) | 45.10(41.51-48.70) |
| | NASNetLarge | 45.64(43.29-47.99) | 41.92(40.35-43.49) | 43.12(39.50-46.74) |
| | Union | 43.92(40.70-47.14) | 38.96(34.63-43.29) | 41.75(38.27-45.24) |
| Accuracy | MobileNetV2 | 77.85(75.10-80.60) | 70.44(67.97-72.91) | 72.84(70.99-74.70) |
| | ResNet101V2 | 74.55(72.55-76.54) | 53.37(45.96-60.78) | 68.27(65.47-71.07) |
| | NASNetLarge | 70.79(69.35-72.24) | 44.46(35.74-53.18) | 71.91(70.08-73.74) |
| | Union | 74.40(72.51-76.29) | 56.09(49.39-62.79) | 71.01(69.44-72.58) |

Form Table 3, we can summarize that the TS-Ensemble model performs much better than the Direct model on external data, and the performances of the TS-Ensemble model on external data are close to the performances of the Direct model on internal data, regarding the metrics of Precision, Recall, F1 and Accuracy. These results indicate that the practical use of the ECDEDL approach for external validation is also quite effective to be able to approximate the internal validation.

## 3.4. Ablation performance of ECDEDL

The visualized metrics of Precision, Recall, F1 and Accuracy at the threshold of different probabilities from 0 to 1 for the TS-Ensemble model on external data and the Direct model respectively on internal data and external data are respectively shown as Figure 5 and Figure 6.

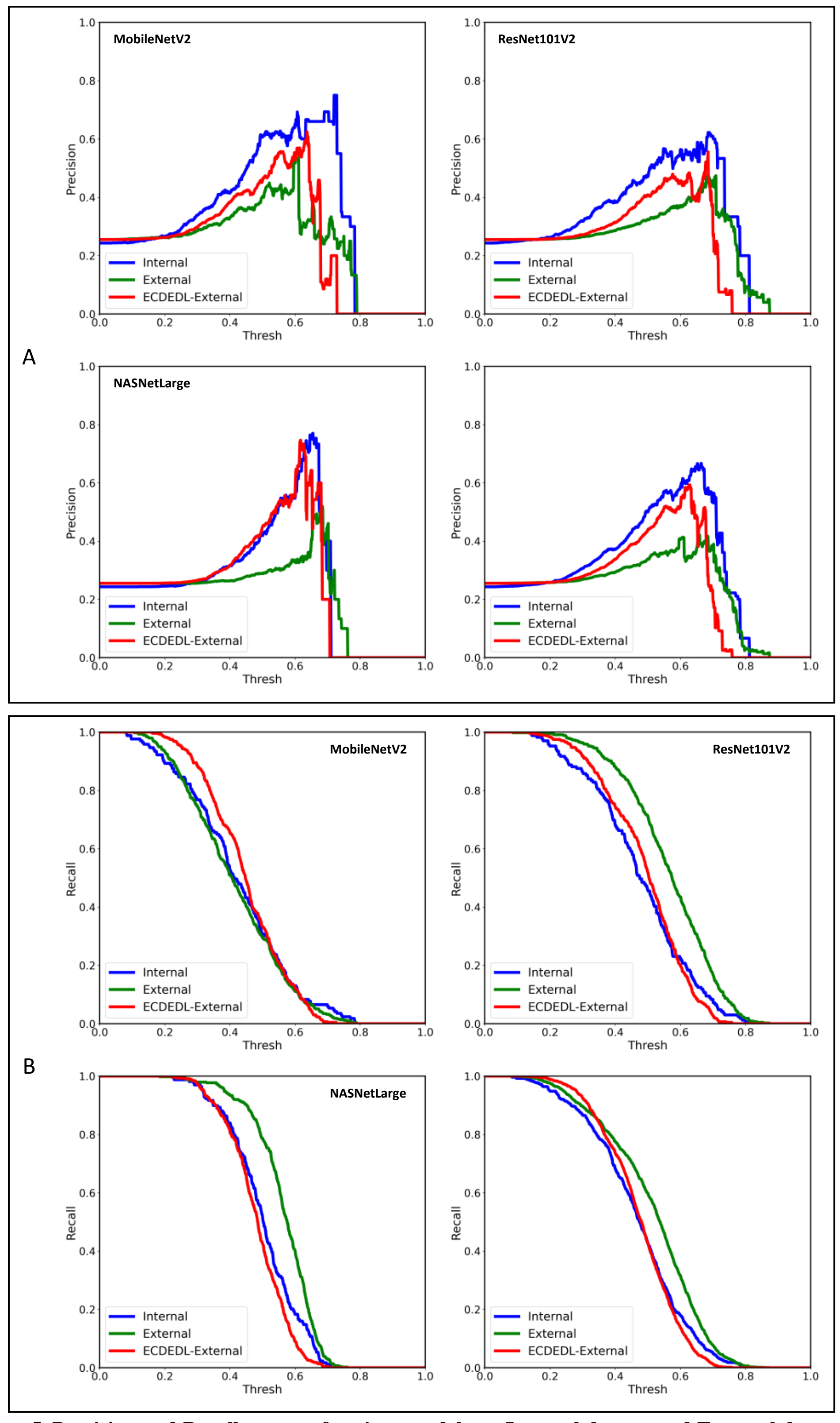

**Figure 5. Precision and Recall curves of various models on Internal dataset and External dataset regarding thresholds of probabilities from 0 to 1. A: Precision curves. B: Recall curves.**

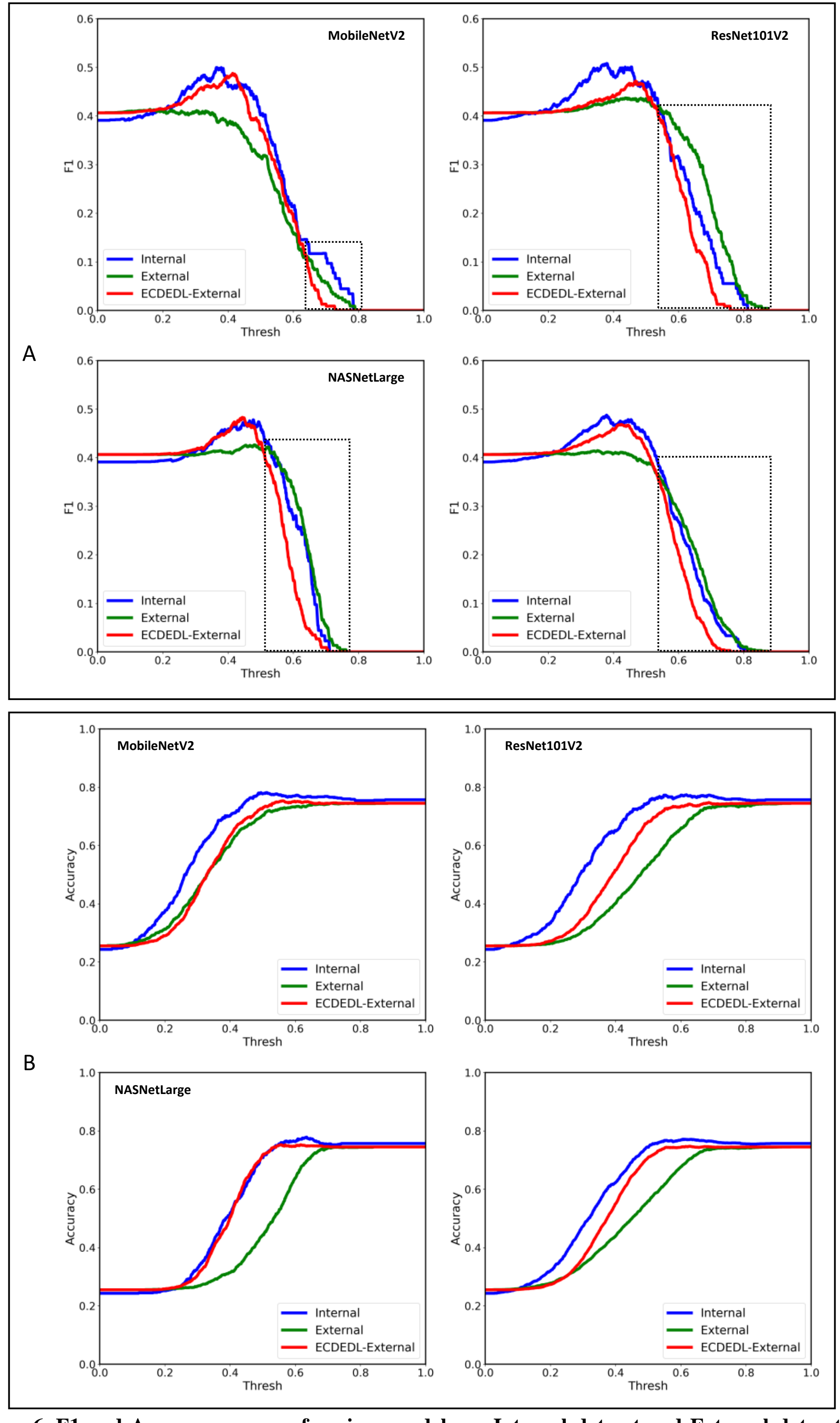

**Figure 6. F1 and Accuracy curves of various models on Internal dataset and External dataset regarding thresholds of probabilities from 0 to 1. A: F1 curves. B: Accuracy curves.**

From Figure 5 and Figure 6, we can observe that the change of the probability threshold will change the performances of different predictive models. Specifically, the overall performances of Accuracy and F1 will generally increase as the threshold of probability value increases, when the threshold of certain probability value is not exceeded. Meanwhile, the overall performances of Accuracy and F1 will generally decrease as the threshold of probability value increases, when the threshold of certain probability value is exceeded.

Notably, from Figure 6-A, we can observe that the F1 curves of ECDEDL-External are generally over the F1 curves of External and are close to the F1 curves of Internal, when the threshold of certain probability value is not exceeded. This indicates that the ECDEDL approach for external validation is quite effective to be able to approximate the internal validation in terms of F1, when the threshold of certain probability value is not exceeded. However, as shown in the dashed bounding boxes in Figure 6-A, the F1 curves of ECDEDL-External are surpassed by the F1 curves of External results, when the threshold of certain probability value is exceeded. This indicates that the proposed ECDEDL approach will be ineffective for external validation in terms of F1, when the threshold of certain probability value is exceeded.

Particularly, from Figure 6-B, we can observe that the Accuracy curves of ECDEDL-External are roughly over the Accuracy curves of External and are close to the Accuracy curves of Internal. This indicates that the ECDEDL approach for external validation is quite effective to be able to approximate the internal validation in terms of Accuracy at most thresholds of probability values, which reflects the stableness of the ECDEDL approach in improving the external validation.

## 5. Conclusion

Directly predicting pCR to NAC in breast cancer from histological images using deep learning has been shown to be a new trend recently [17]. However, it has been a commonly known problem that the AI models constructed for medical prediction have better performances in internal validation than in external validation, which significantly affects the clinical safety of using AI models [23]. Moreover, a recent study [31] indicates that the poor performances in external validation is the primary challenge for deep learning applied to breast cancer imaging. Therefore, it is very meaningful and necessary to investigate advanced approaches for the out-of-distribution problem of external validation in predicting pCR to NAC in breast cancer from histological images.

In this paper, we propose the experts' cognition-driven ensemble deep learning (ECDEDL) approach for external validation and show its effectiveness in predicting pCR to NAC from histological images in breast cancer. Since the cognition of both pathology and AI experts have been taken to construct the predictive model, the proposed ECDEDL approach, to some extent, approximates the working paradigm of a human being which will refer to his various working experiences to make decisions. The proposed ECDEDL approach is more intrinsic than the existing alternative solutions and federated learning for external validation of predicting pCR to NAC in breast cancer from histological images. This property of the proposed ECDEDL approach makes it can fundamentally be combined with existing solutions and federated learning for addressing the problem.

Extensive experimental results and corresponding analysis in this paper primarily indicate: 1) The experts' cognition in the proposed ECDEDL approach is effective with external validation; 2) The overall and practical performances of the proposed ECDEDL approach for external validation are quite effective, numerically approximating the internal validation. These two indications reflect that the proposed ECDEDL approach is closer to the working paradigm of an expert than some existing alternatives solutions which simply ignored the experts' cognition in constructing predictive models for the addressing problem.

The proposed ECDEDL approach still has limitations. The selected experts' cognition ought to have been widely acknowledged because it is used to construct the predictive model in the application of the ECDEDL-based solution for the addressing problem. In fact, the cognition of individual experts that has not been generally accepted can bring personal bias to the final predictive model, leading to biased predictions that can be noxious. Thus, we should try to avoid introducing any expert's cognition that has not been generally accepted in constructing the predictive model of the proposed ECDEDL approach. In addition, although DNN architectures from light weight to complex have been leveraged as the foundation models for investigating the effectiveness of the proposed ECDEDL approach, only three types of architectures were chosen for experiments. This might also lead to biased results regarding to the change of the DNN architecture in implementing the proposed ECDEDL approach. Besides, the number of WSIs collected for external validation was also limited, there might be fluctuations in the experimental results when a larger number of WSIs are available for external validation.

This paper has demonstrated the promising potential of ECDEDL for addressing the out-of-distribution problem of external validation in predicting pCR to NAC in breast cancer from histological images. In future works, it is interesting to explore additional experiments to test the generalizability of the ECDEDL approach across different cancer types or imaging modalities. As the concept behind the proposed ECDEDL approach is fundamentally a new paradigm for artificial intelligence alignment [62], which can be widely leveraged in building predictive models, we particularly look forward to widely testing this paradigm in many other types of cancer-related prediction problems or alternative data preparation strategies [63, 64].

## Recommendations

The proposed experts' cognition-driven ensemble deep learning (ECDEL) approach has the potential to be effective to improve the performance of predictive models for external validation, numerically approximating the internal validation.

## Acknowledgement

We acknowledge Zhongjiu Flash Medical Technology Co., Ltd., Mianyang, China for providing the technical supports for revisions of this paper.

## Funding Support

This work was supported by the 1 • 3 • 5 project for disciplines of excellence (ZYGD18012); the Technological Innovation Project of Chengdu New Industrial Technology Research Institute (2017-CY02 - 00026-GX).

## Ethical Statement

This study does not contain any studies with human or animal subjects performed by any of the authors.

## Conflicts of Interest

The authors declare that they have no conflicts of interest to this work.

## Data Availability Statement

The data that support this work are available upon reasonable request to the corresponding author.

## Author Contribution Statement

**Yongquan Yang:** Conceptualization, Methodology, Software, Validation, Formal analysis, Investigation, Resources, Data curation, Writing - original draft, Visualization, Supervision, Project administration. **Fengling Li:** Conceptualization, Validation, Resources, Data curation, Writing - review & editing. **Yani Wei:** Resources, Writing - review & editing. **Yuanyuan Zhao:** Resources, Writing - review & editing. **Jing Fu:** Resources, Writing - review & editing. **Xiuli Xiao:** Resources, Writing - review & editing. **Hong Bu:** Resources, Writing - review & editing, Supervision, Funding acquisition.

## References

[1] LeCun, Y., Bengio, Y., & Hinton, G. (2015). Deep learning. nature, 521(7553), 436-444. https://doi.org/10.1038/nature14539

[2] Dong, S., Wang, P., & Abbas, K. (2021). A survey on deep learning and its applications. Computer Science Review, 40, 100379. https://doi.org/10.1016/j.cosrev.2021.100379

[3] Han, K., Wang, Y., Chen, H., Chen, X., Guo, J., Liu, Z., ... & Tao, D. (2022). A survey on vision transformer. IEEE transactions on pattern analysis and machine intelligence, 45(1), 87-110. https://doi.org/10.1109/TPAMI.2022.3152247

[4] Khan, S., Naseer, M., Hayat, M., Zamir, S. W., Khan, F. S., & Shah, M. (2022). Transformers in vision: A survey. ACM computing surveys (CSUR), 54(10s), 1-41. https://doi.org/10.1145/3505244

[5] Aggarwal, R., Sounderajah, V., Martin, G., Ting, D. S., Karthikesalingam, A., King, D., ... & Darzi, A. (2021). Diagnostic accuracy of deep learning in medical imaging: a systematic review and meta-analysis. NPJ digital medicine, 4(1), 65. https://doi.org/10.1038/s41746-021-00438-z

[6] Kleppe, A., Skrede, O. J., De Raedt, S., Liestøl, K., Kerr, D. J., & Danielsen, H. E. (2021). Designing deep learning studies in cancer diagnostics. Nature Reviews Cancer, 21(3), 199-211. https://doi.org/10.1038/s41568-020-00327-9

[7] Kuntz, S., Krieghoff-Henning, E., Kather, J. N., Jutzi, T., Höhn, J., Kiehl, L., ... & Brinker, T. J. (2021). Gastrointestinal cancer classification and prognostication from histology using deep learning: Systematic review. European Journal of Cancer, 155, 200-215. https://doi.org/10.1016/j.ejca.2021.07.012

[8] Nam, D., Chapiro, J., Paradis, V., Seraphin, T. P., & Kather, J. N. (2022). Artificial intelligence in liver diseases: Improving diagnostics, prognostics and response prediction. Jhep Reports, 4(4), 100443. https://doi.org/10.1016/j.jhepr.2022.100443

[9] Bleeker, S. E., Moll, H. A., Steyerberg, E. A., Donders, A. R. T., Derksen-Lubsen, G., Grobbee, D. E., & Moons, K. G. M. (2003). External validation is necessary in prediction research:: A clinical example. Journal of clinical epidemiology, 56(9), 826-832. https://doi.org/10.1016/S0895-4356(03)00207-5

[10] Consonni, V., Ballabio, D., & Todeschini, R. (2010). Evaluation of model predictive ability by external validation techniques. Journal of chemometrics, 24(3-4), 194-201. https://doi.org/10.1002/cem.1290

[11] Ramspek, C. L., Jager, K. J., Dekker, F. W., Zoccali, C., & van Diepen, M. (2021). External validation of prognostic models: what, why, how, when and where?. Clinical Kidney Journal, 14(1), 49-58. https://doi.org/10.1093/ckj/sfaa188

[12] Steyerberg, E. W., Bleeker, S. E., Moll, H. A., Grobbee, D. E., & Moons, K. G. (2003). Internal and external validation of predictive models: a simulation study of bias and precision in small samples. Journal of clinical epidemiology, 56(5), 441-447. https://doi.org/10.1016/S0895-4356(03)00047-7

[13] Steyerberg, E. W., & Harrell Jr, F. E. (2016). Prediction models need appropriate internal, internal-external, and external

validation. Journal of clinical epidemiology, 69, 245-247. https://doi.org/10.1016/j.jclinepi.2015.04.005
[14] Derks, M. G., & van de Velde, C. J. (2018). Neoadjuvant chemotherapy in breast cancer: more than just downsizing. The Lancet Oncology, 19(1), 2-3. https://doi.org/10.1016/S1470-2045(17)30914-2
[15] von Minckwitz, G., Blohmer, J. U., Costa, S. D., Denkert, C., Eidtmann, H., Eiermann, W., ... & Loibl, S. (2013). Response-guided neoadjuvant chemotherapy for breast cancer. Journal of Clinical Oncology, 31(29), 3623-3630. https://doi.org/10.1200/JCO.2012.45.0940
[16] Cortazar, P., Zhang, L., Untch, M., Mehta, K., Costantino, J. P., Wolmark, N., ... & Von Minckwitz, G. (2014). Pathological complete response and long-term clinical benefit in breast cancer: the CTNeoBC pooled analysis. The Lancet, 384(9938), 164-172. https://doi.org/10.1016/S0140-6736(13)62422-8
[17] Echle, A., Rindtorff, N. T., Brinker, T. J., Luedde, T., Pearson, A. T., & Kather, J. N. (2021). Deep learning in cancer pathology: a new generation of clinical biomarkers. British journal of cancer, 124(4), 686-696. https://doi.org/10.1038/s41416-020-01122-x
[18] Beck, A. H., Sangoi, A. R., Leung, S., Marinelli, R. J., Nielsen, T. O., Van De Vijver, M. J., ... & Koller, D. (2011). Systematic analysis of breast cancer morphology uncovers stromal features associated with survival. Science translational medicine, 3(108), 108ra113. https://doi.org/10.1126/scitranslmed.3002564
[19] Bhargava, H. K., Leo, P., Elliott, R., Janowczyk, A., Whitney, J., Gupta, S., ... & Madabhushi, A. (2020). Computationally derived image signature of stromal morphology is prognostic of prostate cancer recurrence following prostatectomy in African American patients. Clinical Cancer Research, 26(8), 1915-1923. https://doi.org/10.1158/1078-0432.CCR-19-2659
[20] Kather, J. N., Krisam, J., Charoentong, P., Luedde, T., Herpel, E., Weis, C. A., ... & Halama, N. (2019). Predicting survival from colorectal cancer histology slides using deep learning: A retrospective multicenter study. PLoS medicine, 16(1), e1002730. https://doi.org/10.1371/journal.pmed.1002730
[21] Li, B., Li, F., Liu, Z., Xu, F., Ye, G., Li, W., ... & Tian, J. (2022). Deep learning with biopsy whole slide images for pretreatment prediction of pathological complete response to neoadjuvant chemotherapy in breast cancer: A multicenter study. The Breast, 66, 183-190. https://doi.org/10.1016/j.breast.2022.10.004
[22] Zhang, F., Yao, S., Li, Z., Liang, C., Zhao, K., Huang, Y., ... & Liu, Z. (2020). Predicting treatment response to neoadjuvant chemoradiotherapy in local advanced rectal cancer by biopsy digital pathology image features. Clinical and translational medicine, 10(2), e110. https://doi.org/10.1002/ctm2.110
[23] Challen, R., Denny, J., Pitt, M., Gompels, L., Edwards, T., & Tsaneva-Atanasova, K. (2019). Artificial intelligence, bias and clinical safety. BMJ quality & safety, 28(3), 231-237. https://doi.org/10.1136/bmjqs-2018-008370
[24] Cortes, C., Mohri, M., Riley, M., & Rostamizadeh, A. (2008). Sample selection bias correction theory. In International conference on algorithmic learning theory, 38-53. https://doi.org/10.1007/978-3-540-87987-9_8
[25] Huang, J., Gretton, A., Borgwardt, K., Schölkopf, B., & Smola, A. (2006). Correcting sample selection bias by unlabeled data. In Advances in Neural Information Processing Systems 19: Proceedings of the 2006 Conference, 601-608. https://doi.org/10.7551/mitpress/7503.003.0080
[26] Teney, D., Abbasnejad, E., Kafle, K., Shrestha, R., Kanan, C., & Van Den Hengel, A. (2020). On the value of out-of-distribution testing: An example of goodhart's law. In Advances in neural information processing systems, 33, 407-417.
[27] Guan, H., & Liu, M. (2021). Domain adaptation for medical image analysis: a survey. IEEE Transactions on Biomedical Engineering, 69(3), 1173-1185. https://doi.org/10.1109/TBME.2021.3117407
[28] Patel, V. M., Gopalan, R., Li, R., & Chellappa, R. (2015). Visual domain adaptation: A survey of recent advances. IEEE signal processing magazine, 32(3), 53-69. https://doi.org/10.1109/MSP.2014.2347059
[29] Wang, J., Lan, C., Liu, C., Ouyang, Y., Qin, T., Lu, W., ... & Philip, S. Y. (2022). Generalizing to unseen domains: A survey on domain generalization. IEEE transactions on knowledge and data engineering, 35(8), 8052-8072. https://doi.org/10.1109/TKDE.2022.3178128
[30] Zhou, K., Liu, Z., Qiao, Y., Xiang, T., & Loy, C. C. (2022). Domain generalization: A survey. IEEE Transactions on Pattern Analysis and Machine Intelligence, 45(4), 4396-4415. https://doi.org/10.1109/TPAMI.2022.3195549
[31] Luo, L., Wang, X., Lin, Y., Ma, X., Tan, A., Chan, R., ... & Chen, H. (2024). Deep learning in breast cancer imaging: A decade of progress and future directions. IEEE Reviews in Biomedical Engineering, 1-20. https://doi.org/10.1109/RBME.2024.3357877
[32] Faryna, K., van der Laak, J., & Litjens, G. (2021). Tailoring automated data augmentation to H&E-stained histopathology. In Proceedings of Machine Learning Research, 1-11.
[33] Tellez, D., Balkenhol, M., Otte-Höller, I., van de Loo, R., Vogels, R., Bult, P., ... & Ciompi, F. (2018). Whole-slide mitosis detection in H&E breast histology using PHH3 as a reference to train distilled stain-invariant convolutional networks. IEEE transactions on medical imaging, 37(9), 2126-2136. https://doi.org/10.1109/TMI.2018.2820199
[34] de Haan, K., Zhang, Y., Zuckerman, J. E., Liu, T., Sisk, A. E., Diaz, M. F., ... & Ozcan, A. (2021). Deep learning-based transformation of H&E stained tissues into special stains. Nature communications, 12(1), 1-13. https://doi.org/10.1038/s41467-021-25221-2
[35] Mahmood, F., Borders, D., Chen, R. J., McKay, G. N., Salimian, K. J., Baras, A., & Durr, N. J. (2019). Deep adversarial training for multi-organ nuclei segmentation in histopathology images. IEEE transactions on medical imaging, 39(11), 3257-3267. https://doi.org/10.1109/TMI.2019.2927182
[36] Alirezazadeh, P., Hejrati, B., Monsef-Esfahani, A., & Fathi, A. (2018). Representation learning-based unsupervised domain adaptation for classification of breast cancer histopathology images. Biocybernetics and Biomedical Engineering, 38(3), 671-

683. https://doi.org/10.1016/j.bbe.2018.04.008
[37] Lafarge, M. W., Pluim, J. P., Eppenhof, K. A., Moeskops, P., & Veta, M. (2017). Domain-adversarial neural networks to address the appearance variability of histopathology images. In Deep Learning in Medical Image Analysis and Multimodal Learning for Clinical Decision Support, 83-91. https://doi.org/10.1007/978-3-319-67558-9_10
[38] Zhang, Y., Chen, H., Wei, Y., Zhao, P., Cao, J., Fan, X., ... & Huang, J. (2019, October). From whole slide imaging to microscopy: Deep microscopy adaptation network for histopathology cancer image classification. In International conference on medical image computing and computer-assisted intervention, 360-368. https://doi.org/10.1007/978-3-030-32239-7_40
[39] Li, C., Lin, X., Mao, Y., Lin, W., Qi, Q., Ding, X., ... & Yu, Y. (2022). Domain generalization on medical imaging classification using episodic training with task augmentation. Computers in biology and medicine, 141, 105144. https://doi.org/10.1016/j.compbiomed.2021.105144
[40] Li, H., Wang, Y., Wan, R., Wang, S., Li, T. Q., & Kot, A. (2020). Domain generalization for medical imaging classification with linear-dependency regularization. In Advances in neural information processing systems, 33, 3118-3129.
[41] Liu, Q., Chen, C., Qin, J., Dou, Q., & Heng, P. A. (2021). Feddg: Federated domain generalization on medical image segmentation via episodic learning in continuous frequency space. In Proceedings of the IEEE/CVF conference on computer vision and pattern recognition, 1013-1023. https://doi.org/10.1109/CVPR46437.2021.00107
[42] Ouyang, C., Chen, C., Li, S., Li, Z., Qin, C., Bai, W., & Rueckert, D. (2022). Causality-inspired single-source domain generalization for medical image segmentation. IEEE Transactions on Medical Imaging, 42(4), 1095-1106. https://doi.org/10.1109/TMI.2022.3224067
[43] Ogier du Terrail, J., Leopold, A., Joly, C., Béguier, C., Andreux, M., Maussion, C., ... & Heudel, P. E. (2023). Federated learning for predicting histological response to neoadjuvant chemotherapy in triple-negative breast cancer. Nature medicine, 29(1), 135-146. https://doi.org/10.1038/s41591-022-02155-w
[44] Sarma, K. V., Harmon, S., Sanford, T., Roth, H. R., Xu, Z., Tetreault, J., ... & Arnold, C. W. (2021). Federated learning improves site performance in multicenter deep learning without data sharing. Journal of the American Medical Informatics Association, 28(6), 1259-1264. https://doi.org/10.1093/jamia/ocaa341
[45] Banabilah, S., Aloqaily, M., Alsayed, E., Malik, N., & Jararweh, Y. (2022). Federated learning review: Fundamentals, enabling technologies, and future applications. Information processing & management, 59(6), 103061. https://doi.org/10.1016/j.ipm.2022.103061
[46] Nguyen, D. C., Pham, Q. V., Pathirana, P. N., Ding, M., Seneviratne, A., Lin, Z., ... & Hwang, W. J. (2022). Federated learning for smart healthcare: A survey. ACM Computing Surveys (Csur), 55(3), 1-37. https://doi.org/10.1145/3501296
[47] Zhang, C., Xie, Y., Bai, H., Yu, B., Li, W., & Gao, Y. (2021). A survey on federated learning. Knowledge-Based Systems, 216, 106775. https://doi.org/10.1016/j.knosys.2021.106775
[48] Yang, Y., Lv, H., & Chen, N. (2023). A survey on ensemble learning under the era of deep learning. Artificial Intelligence Review, 56(6), 5545-5589. https://doi.org/10.1007/s10462-022-10283-5
[49] Li, F., Yang, Y., Wei, Y., He, P., Chen, J., Zheng, Z., & Bu, H. (2021). Deep learning-based predictive biomarker of pathological complete response to neoadjuvant chemotherapy from histological images in breast cancer. Journal of translational medicine, 19, 1-13. https://doi.org/10.1186/s12967-021-03020-z
[50] Li, F., Yang, Y., Wei, Y., Zhao, Y., Fu, J., Xiao, X., ... & Bu, H. (2022). Predicting neoadjuvant chemotherapy benefit using deep learning from stromal histology in breast cancer. NPJ Breast Cancer, 8(1), 124. https://doi.org/10.1038/s41523-022-00491-1
[51] Dietterich, T. G. (2000). Ensemble methods in machine learning. In International workshop on multiple classifier systems, 1-15. https://doi.org/10.1007/3-540-45014-9_1
[52] Zhou, Z. H. (2012). Ensemble methods: foundations and algorithms. USA: CRC press.
[53] Dubey, A., Ramanathan, V., Pentland, A., & Mahajan, D. (2021). Adaptive methods for real-world domain generalization. In Proceedings of the IEEE/CVF Conference on Computer Vision and Pattern Recognition, 14340-14349. https://doi.org/10.1109/CVPR46437.2021.01411
[54] Wu, G., & Gong, S. (2021). Collaborative optimization and aggregation for decentralized domain generalization and adaptation. In Proceedings of the IEEE/CVF International Conference on Computer Vision, 6484-6493. https://doi.org/10.1109/ICCV48922.2021.00642
[55] Zhou, K., Yang, Y., Qiao, Y., & Xiang, T. (2021). Domain adaptive ensemble learning. IEEE Transactions on Image Processing, 30, 8008-8018. https://doi.org/10.1109/TIP.2021.3112012
[56] Yang, Y., Li, F., Wei, Y., Chen, J., Chen, N., Alobaidi, M. H., & Bu, H. (2024). One-step abductive multi-target learning with diverse noisy samples and its application to tumour segmentation for breast cancer. Expert Systems with Applications, 251, 123923. https://doi.org/10.1016/j.eswa.2024.123923
[57] Khan, A., Sohail, A., Zahoora, U., & Qureshi, A. S. (2020). A survey of the recent architectures of deep convolutional neural networks. Artificial intelligence review, 53, 5455-5516. https://doi.org/10.1007/s10462-020-09825-6
[58] Sandler, M., Howard, A., Zhu, M., Zhmoginov, A., & Chen, L. C. (2018). Mobilenetv2: Inverted residuals and linear bottlenecks. In Proceedings of the IEEE conference on computer vision and pattern recognition, 4510-4520. https://doi.org/10.1109/CVPR.2018.00474
[59] He, K., Zhang, X., Ren, S., & Sun, J. (2016). Identity mappings in deep residual networks. In Computer Vision–ECCV 2016: 14th European Conference, 630-645. https://doi.org/10.1007/978-3-319-46493-0_38
[60] Zoph, B., Vasudevan, V., Shlens, J., & Le, Q. V. (2018). Learning transferable architectures for scalable image recognition.

In Proceedings of the IEEE conference on computer vision and pattern recognition, 8697-8710. https://doi.org/10.1109/CVPR.2018.00907
[61] Theodoridis, S. (2015). Stochastic gradient descent. In S. Theodoridis (Ed.), Machine Learning: A Bayesian and Optimization Perspective (pp. 161-231).Academic Press. https://doi.org/10.1016/B978-0-12-801522-3.00005-7
[62] Yang, Y. (2023). Discovering Scientific Paradigms for Artificial Intelligence Alignment. http://dx.doi.org/10.13140/RG.2.2.15945.52320
[63] Jain, S., Naicker, D., Raj, R., Patel, V., Hu, Y. C., Srinivasan, K., & Jen, C. P. (2023). Computational intelligence in cancer diagnostics: a contemporary review of smart phone apps, current problems, and future research potentials. Diagnostics, 13(9), 1563. https://doi.org/10.3390/diagnostics13091563
[64] Murthy, N. S., & Bethala, C. (2023). Review paper on research direction towards cancer prediction and prognosis using machine learning and deep learning models. Journal of Ambient Intelligence and Humanized Computing, 14(5), 5595-5613. https://doi.org/10.1007/s12652-021-03147-3